\documentclass[prx,twocolumn, amsmath,amssymb,
reprint,%
author-year,%
author-numerical,%
dvipdfmx
]{revtex4}

\usepackage{graphicx}
\usepackage{bm}
\usepackage{color}

\begin{document}


\title{Lattice stability of ordered Au-Cu alloys in the warm dense matter regime}

\author{Shota Ono}
\email{shota_o@gifu-u.ac.jp}
\author{Daigo Kobayashi}
\affiliation{Department of Electrical, Electronic and Computer Engineering, Gifu University, Gifu 501-1193, Japan}

\begin{abstract}
In the warm dense regime, where the electron temperature is increased to the same order of the Fermi temperature, the dynamical stability of elemental metals depends on its electronic band structure as well as its crystal structure. It has been known that phonon hardening occurs due to an enhanced internal pressure caused by electron excitations as in close-packed simple metals, whereas phonon softening occurs at a specific point in the Brillouin zone as in body-centered cubic metals. Here, we investigate the dynamical stability of binary ordered alloys (Au and Cu) in the L1$_0$ and L1$_2$ structures. By performing first principles calculations on phonon dispersions, we demonstrate that warm dense Au-Cu systems show phonon hardening behavior except the lowest frequency phonon mode at point R of AuCu in the L1$_0$ structure. We show that when a phonon mode is stabilized by long-range interatomic interactions at ambient condition, such a phonon will be destabilized by the short-range nature of the warm dense matters.
\end{abstract}

\maketitle

\section{Introduction} Lattice dynamics in metals is determined by the interplay between the direct ion-ion and the indirect ion-electron-ion interactions. In the warm dense matter (WDM) regime realized by an incident of ultrashort laser pulse to metals, the indirect intereaction is strongly modulated due to electron excitations characterized by the electron temperature $T_{\rm e}$ that is the same order of the Fermi temperature. By using the finite-temperature density-functional theory (DFT) \cite{mermin}, the phonon hardening has been predicted for elemental metals in the close-packed (i.e., face-centered cubic (fcc) and hexagonal close-packed) structures \cite{recoules,bottin,kabeer,giret,minakov,yan,harbour,ono2019}. For noble metals, this is because the excitation of $d$ electrons weakens the electron screening, in turn, yielding an increase in the internal pressure \cite{recoules,bottin}. For free electron metals with no $d$ bands near the Fermi level, the internal pressure increases with the electron kinetic energy, leading to the phonon hardening again \cite{bottin}. Within the free-electron approximation using the central potential model, this is understood by an increase in the force constants of the first nearest neighbor (NN) sites \cite{ono2019}. In contrast, the phonon softening has been predicted for the body-centered cubic (bcc) structure \cite{giret,yan,harbour,ono2019,ono2020}. This is intrinsic to the crystal structure symmetry: in the expression for the phonon frequency at point N in the Brillouin zone, a relatively small contribution from the force constants of the first NN sites is present \cite{ono2019,ono2020}. These studies imply that the phonon hardening or softening properties in WDM are determined by a combined effect of the increase in the internal pressure (weakened screening) and the crystal structure symmetry. In order to study this interplay systematically, it is useful to focus on ordered alloys that can have many crystal structures in thermal equilibrium and phase diagram as a function of ambient temperature and composition. 

\begin{figure}[bb]
\center
\includegraphics[scale=0.55]{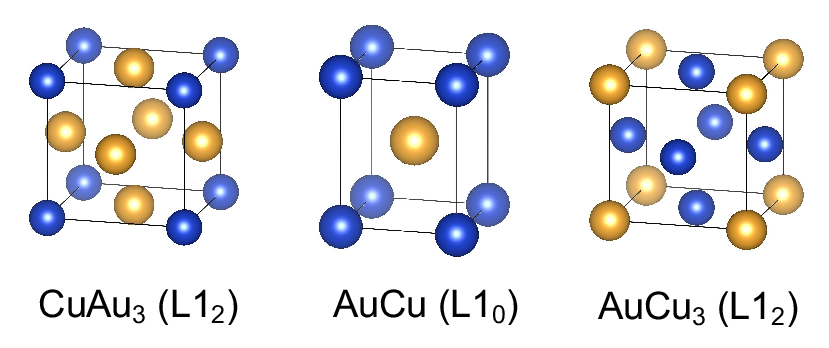}
\caption{Crystal structure of CuAu$_3$ (L1$_2$), AuCu (L1$_0$), and AuCu$_3$ (L1$_2$).  } \label{fig1} 
\end{figure}

The ordered phases of Au-Cu alloys have been well known as described in the standard textbooks \cite{kittel}: AuCu in the L1$_0$ structure (the distorted fcc structure along the $c$ axis) and AuCu$_3$ and CuAu$_3$ in the L1$_2$ structure \cite{okamoto}, as shown in Fig.~\ref{fig1}. In this paper, we study the lattice dynamics of ordered Au-Cu alloys in the WDM regime. We demonstrate that the warm dense Au-Cu alloy in the L1$_2$ structure shows the phonon hardening, whereas that in the L1$_0$ structure is unstable against the lowest frequency phonons at point R. The former can be explained by the weakening of the electron screening effect. To explain the latter, we show that the corresponding phonon mode at point R is stabilized by long-range interatomic interactions at ambient temperature due to the crystal structure symmetry and that the short-range nature of the warm dense matters destabilizes such a phonon mode.

The origin of the phonon softening in L1$_0$ AuCu alloy is different from that in warm dense diamond and silicon. For the latter, the phonon softening and lattice instabilities have been understood by the strengthening of the anti-bonding character between atoms created by the electron transition from the valence to the conduction bands, reducing the size of force constants significantly \cite{stampfli}. Recently, Yan et al. have studied the dynamical stability of warm dense tantalum nitrides (TaN) in various structures \cite{yan2020} using finite-temperature DFT. They have shown that under electronic excitations the unstable cubic $\delta$-phase becomes stable, whereas the hexagonal phases show the phonon softening. They have explained it with respect to the covalent bonding between atoms in more complicated way. 

This paper is organized as follows. Section \ref{sec:comp} describes computational details for the total energy and phonon dispersion calculations based on DFT and density-functional perturbation theory (DFPT). Section \ref{sec:formulae} briefly explains the theory of lattice dynamics to provide analytical expressions for phonon frequencies at point R of AuCu in the L1$_0$ structure. Section \ref{sec:results} first investigates the equilibrium and phonon properties of Au-Cu alloys by using different exchange-correlation functionals. The phonon dispersions of Au-Cu alloys in the WDM regime is next presented, followed by a consideration of the electron screening that explains the phonon hardening behaviors. The analytical expressions of phonon frequencies derived in Sec.~\ref{sec:formulae} will be systematically used to explain the phonon softening at point R of L1$_0$ AuCu in the WDM regime. Also, nonthermal solid-to-solid transformation of L1$_0$ AuCu is discussed. We conclude the paper in Sec.~\ref{sec:conclusion}.


\section{Theory} 
\subsection{Computational details} 
\label{sec:comp}
We calculate the total energy and the phonon dispersions of Au-Cu alloys based on DFT and DFPT \cite{dfpt} implemented in \texttt{Quantum ESPRESSO} (\texttt{QE}) code \cite{qe}. In order to treat the exchange-correlation energy, we use the Perdew-Zunger (PZ) \cite{pz} functional of the local density approximation and the Perdew-Burke-Ernzerhof (PBE) \cite{pbe} functionals of the generalized gradient approximation. We use the ultrasoft pseudopotentials of Au.$xc$-n-rrkjus\_psl.1.0.0.UPF and Cu.$xc$-dn-rrkjus\_psl.1.0.0.UPF $(xc={\rm pz \ or \ pbe})$ that are provided in \texttt{pslibrary.1.0.0} \cite{dalcorso}. The cutoff energies for the wavefunction and the charge density are 80 Ry and 800 Ry, respectively. The self-consistent field and phonon dispersion calculations are performed by using 16$\times$16$\times$16 $k$ grid and 4$\times$4$\times$4 $q$ grid (3$\times$3$\times$3 $q$ grid for L1$_2$) \cite{MK}, respectively, which are enough to study the dynamical stability of warm dense alloys. The total energy and forces are converged within $10^{-5}$ Ry and $10^{-4}$ a.u., respectively in geometry optimization. For phonon calculations, the threshold parameter for self-consistency (tr2\_ph) is set to $10^{-14}$. The Marzari-Vanderbilt smearing \cite{smearing} with a broadening of $\sigma=0.02$ Ry (0.27 eV) is used for the geometry optimization, while the Fermi-Dirac type smearing is used for the WDM regime (the electron temperature $T_{\rm e}$ in units of eV and the Boltzmann constant $k_{\rm B}=1$). The tetrahedron method is used to calculate the electron density-of-states (DOS) \cite{tetra}. For the WDM regime, the number of electronic bands to be calculated is increased to more than three times larger the sizes in the ground state calculations in order to take into account the occupation of electronic states far above the Fermi level. Imaginary phonon energies $\hbar\omega$ ($\hbar$ the Planck constant and $\omega$ the phonon frequency) are represented by negative energies below. 


\subsection{Analytical expressions for phonon frequencies}
\label{sec:formulae}

Below is a brief description of the basic concepts for the lattice dynamics. We derive analytical expressions for the phonon frequencies at point R in order to study the instability of L1$_0$ AuCu in detail. The atomic motion in a crystal is regulated by \cite{maradudin}
\begin{eqnarray}
 M_\kappa \frac{d^2 u_{\kappa\alpha}(\bm{R}_i)}{dt^2} 
 = - \sum_{\kappa' \beta j} D_{\alpha\beta}^{\kappa\kappa'}(\bm{R}_i,\bm{R}_j)
u_{\kappa'\beta}(\bm{R}_j),
 \label{eq:newton}
\end{eqnarray}
where $u_{\kappa\alpha}(\bm{R}_i)$ is the displacement components for the atom $\kappa$ ($=1$ and 2 for Au and Cu, respectively) along the direction of $\alpha \ (=x,y,z)$ in the unit cell characterized by the lattice vector $\bm{R}_i=(R_{ix},R_{iy},R_{iz})$ expanded by the primitive vectors $\bm{a}_1=(a,0,0), \bm{a}_2=(0,a,0)$, and $\bm{a}_3=(0,0,c)$. $M_\kappa$ is the mass of the atom $\kappa$ and $D_{\alpha\beta}^{\kappa\kappa'}(\bm{R}_i,\bm{R}_j)$ is the force constant matrix defined as
\begin{eqnarray}
D_{\alpha\beta}^{\kappa\kappa'}(\bm{R}_i,\bm{R}_j)
= \frac{\partial^2 V}{\partial X_{\kappa\alpha}(\bm{R}_i) \partial X_{\kappa'\beta}(\bm{R}_j) }\Big\vert_0,
\end{eqnarray}
where $X_{\kappa\alpha}(\bm{R}_i) = R_{i\alpha} + \tau_{\kappa\alpha}$ and $\tau_{\kappa\alpha}$ is the $\alpha$-component of the basis vector for $\kappa$: $(\tau_{1x},\tau_{1y},\tau_{1z})=(0,0,0)$ and $(\tau_{2x},\tau_{2y},\tau_{2z})=(a/2,a/2,c/2)$. Assuming the plane wave solution with the wavevector $\bm{q}$, one obtains the eigenvalue equation
\begin{eqnarray}
\omega^2 \epsilon_{\kappa\alpha}(\bm{q}) = 
\sum_{\kappa' \beta} 
\tilde{D}_{\alpha\beta}^{\kappa\kappa'}(\bm{q})
 \epsilon_{\kappa'\beta}(\bm{q}),
  \label{eq:eig}
\end{eqnarray}
where $\epsilon_{\kappa\alpha}$ is the $\alpha$-component of the polarization vector of the atom $\kappa$ and, using the translational symmetry of the crystal, the dynamical matrix $\tilde{D}$ is given by
\begin{eqnarray}
\tilde{D}_{\alpha\beta}^{\kappa\kappa'}(\bm{q})
 = \frac{1}{\sqrt{M_\kappa M_{\kappa'}}}
 \sum_{j} D_{\alpha\beta}^{\kappa\kappa'}(\bm{R}_j) 
 e^{-i\bm{q}\cdot \bm{R}_j},
 \label{eq:dyn}
\end{eqnarray}
where we used the notation of $D_{\alpha\beta}^{\kappa\kappa'}(\bm{R}_j) = D_{\alpha\beta}^{\kappa\kappa'}(\bm{R}_j,\bm{0})$. Also, due to the translational symmetry, the acoustic sum rule holds: $\sum_{\kappa' j} D_{\beta\alpha}^{\kappa'\kappa}(\bm{R}_j) =0$.

We consider the force constants up to the fifth-order NN sites of L1$_0$ AuCu: for the Au atom at the origin, the position of the first, second, third, fourth, and fifth NNs are Cu$(a/2,a/2,c/2)$, Au$(a,0,0)$, Au$(0,0,c)$, Au$(a,0,c)$, and Cu$(3a/2,a/2,c/2)$, respectively, and these equivalent sites. The $6\times 6$ dynamical matrix is then expressed as
\begin{eqnarray}
\tilde{D}(\bm{q}) = 
 \left(
 \begin{array}{cc}
 \tilde{D}^{11}(\bm{q}) & \tilde{D}^{12}(\bm{q})  \\
 \tilde{D}^{21}(\bm{q}) & \tilde{D}^{22}(\bm{q})  \\
\end{array}
\right),
\end{eqnarray}
where the $3\times 3$ diagonal matrix is given by 
\begin{eqnarray}
\tilde{D}^{\kappa\kappa}(\bm{q})=
 \left(
 \begin{array}{ccc}
 \tilde{D}_{xx}^{\kappa\kappa}(\bm{q}) & 0 & 0  \\
 0 & \tilde{D}_{yy}^{\kappa\kappa}(\bm{q}) & 0  \\
 0 & 0 & \tilde{D}_{zz}^{\kappa\kappa}(\bm{q})   \\
\end{array}
\right)
\end{eqnarray}
and the $3\times 3$ off-diagonal matrix for $\kappa\ne\kappa'$ is given by
\begin{eqnarray}
\tilde{D}^{\kappa\kappa'}(\bm{q})=
 \left(
 \begin{array}{ccc}
 0 & 0 & 0  \\
 0 & 0 & \tilde{D}_{yz}^{\kappa\kappa'}(\bm{q})   \\
 0 & \tilde{D}_{zy}^{\kappa\kappa'}(\bm{q}) & 0  \\
\end{array}
\right). 
\end{eqnarray}
By solving Eq.~(\ref{eq:eig}), one obtains the phonon energies at point R in ascending order
\begin{eqnarray}
\omega_{\rm I}^{2} &=& \frac{1}{2}\left[ \tilde{D}_{zz}^{11}(\bm{q})+\tilde{D}_{yy}^{22}(\bm{q}) 
- \tilde{C}_{zy}(\bm{q}) \right],
\nonumber\\
\omega_{\rm II}^{2} &=& \tilde{D}_{xx}^{11}(\bm{q}),
\nonumber\\
\omega_{\rm III}^{2} &=& \frac{1}{2}\left[\tilde{D}_{yy}^{11}(\bm{q})+\tilde{D}_{zz}^{22}(\bm{q}) 
- \tilde{C}_{yz}(\bm{q}) \right],
\nonumber\\
\omega_{\rm IV}^{2} &=& \tilde{D}_{xx}^{22}(\bm{q}),
\nonumber\\
\omega_{\rm V}^{2} &=& \frac{1}{2}\left[\tilde{D}_{zz}^{11}(\bm{q})+\tilde{D}_{yy}^{22}(\bm{q}) 
+ \tilde{C}_{zy}(\bm{q}) \right],
\nonumber\\
\omega_{\rm VI}^{2} &=& \frac{1}{2}\left[\tilde{D}_{yy}^{11}(\bm{q}) +\tilde{D}_{zz}^{22}(\bm{q}) 
+ \tilde{C}_{yz}(\bm{q}) \right]
\label{eq:omegaI-VI}
\end{eqnarray}
with 
\begin{eqnarray}
\tilde{C}_{\alpha\beta}(\bm{q}) &=& \sqrt{\left[\tilde{D}_{\alpha\alpha}^{11}(\bm{q})
-\tilde{D}_{\beta\beta}^{22}(\bm{q})\right]^2
+4\left[\tilde{D}_{\alpha\beta}^{12}(\bm{q})\right]^2}.
\nonumber\\
\label{eq:Cq}
\end{eqnarray}
Based on these expressions for $\omega_{k} \ (k={\rm I} \cdots {\rm VI})$, we interpret the lattice vibrations as follows: the vibration of the Au (Cu) atoms polarized to the $x$ direction propagates along the $\bm{q}=(0,\pi/a,\pi/c)$ direction with the frequency $\omega_{\rm II}$ ($\omega_{\rm IV}$); the vibration of the Au atoms polarized to the $z$ direction is coupled with the vibration of the Cu atoms polarized to the $y$ direction, yielding $\omega_{\rm I}$ and $\omega_{\rm V}$; and the vibration of the Au atoms polarized to the $y$ direction is coupled with the vibration of the Cu atoms polarized to the $z$ direction, yielding $\omega_{\rm III}$ and $\omega_{\rm VI}$. Due to the mode mixing, a frequency gap must be present between $\omega_{\rm I}$ and $\omega_{\rm V}$ ($\omega_{\rm III}$ and $\omega_{\rm VI}$). Negative value in the expression of $\omega_{\rm I}^{2}$ (i.e., $- \tilde{C}_{zy}(\bm{q})$) will be responsible for the instability of L1$_0$ AuCu in the WDM regime. The expressions of $\tilde{D}_{\alpha\beta}^{\kappa\kappa'}(\bm{q})$ as a function of $D_{\alpha\beta}^{\kappa\kappa'}(\bm{R}_j)$ are provided in the Appendix \ref{app}. The values of $D_{\alpha\beta}^{\kappa\kappa'}(\bm{R}_j)$ are extracted from DFPT calculations described in Sec.~\ref{sec:comp}. 



\begin{table}[bb]
\begin{center}
\caption{Lattice parameters ($a$ and $c$ in units of \AA) of Au-Cu alloys for different functionals. The experimental values are extracted from Ref.~\cite{okamoto}. }
{
\begin{tabular}{lccc}\hline\hline
   \hspace{10mm}  & \hspace{5mm} PZ \hspace{5mm}    & \hspace{5mm} PBE \hspace{5mm}  &  Experiment       \\  \hline
Au & 4.06 & 4.16 &  4.08 \\
CuAu$_3$ & 3.95 & 4.05  & 3.95 \\
AuCu $(a)$ & 2.80 & 2.88  & 2.85 \\
AuCu $(c)$ & 3.54 & 3.64  &  3.67 \\
AuCu$_3$ & 3.68 & 3.79  & 3.74 \\
Cu & 3.53 & 3.63 & 3.62 \\
MAE & 0.060 & 0.051 & \\
\hline\hline
\end{tabular}
}
\label{table1}
\end{center}
\end{table}

\begin{figure}[tt]
\center
\includegraphics[scale=0.35]{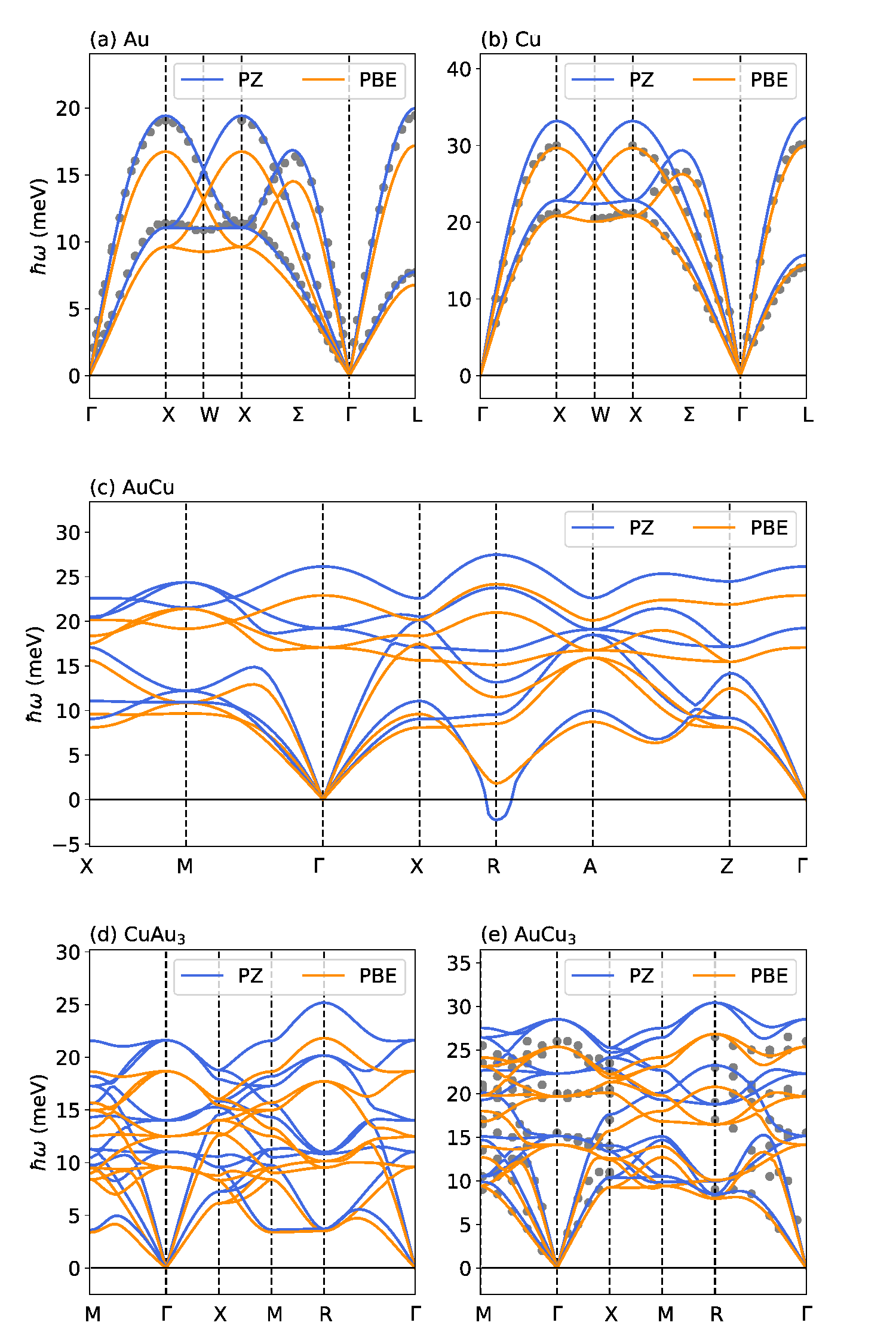}
\caption{The phonon dispersions of (a) Au, (b) Cu, (c) AuCu, (d) CuAu$_3$, and (e) AuCu$_3$ calculated with the PZ and PBE functionals. The experimental data (circles) for (a), (b), and (e) are taken from Refs.~\cite{lynn}, \cite{nicklow}, and \cite{katano}, respectively. } \label{fig_functional} 
\end{figure}

\section{Results and discussion}
\label{sec:results}
\subsection{Functional dependence}
\label{sec:functional}

It has been known that the cohesive energies and bulk moduli of alloys may be sensitive to the exchange-correlation functionals used \cite{zhang,isaacs,ruzsinszky2019,ruzsinszky2020}. In general, the PBE functional underestimates those values for weakly bonded systems such as Au-Cu alloys. Such a discrepancy can be improved by using the hybrid exchange-correlation functional \cite{zhang} or the random phase approximation \cite{ruzsinszky2019}. For more accuracy, it may be better to use the latter approaches \cite{zhang,ruzsinszky2019}. However, we use the standard functionals described above to reduce the computational costs in phonon dispersion calculations.

For studying numerical accuracy in the present calculations, we list the optimized lattice parameters $a$ and $c$ (only for L1$_0$) predicted by using PZ and PBE functionals in Table~\ref{table1}. These values obtained at $T_{\rm e}=0$ agree with the experimental values \cite{okamoto}: The mean absolute error (MAE) is estimated to be less than 0.1 \AA \ for both cases. For Au-rich phases the PZ results are in good agreement with the experimental lattice constants, while for Cu-rich phases the PBE results are better.

Figure \ref{fig_functional} shows the phonon dispersions along the symmetry lines for Au-Cu systems: (a) fcc Au, (b) fcc Cu, (c) L1$_0$ AuCu, (d) L1$_2$ CuAu$_3$, and (e) L1$_2$ AuCu$_3$. Overall, the values of $\omega$s calculated by using the PBE functional seem to be smaller than those calculated by using the PZ. For simple metals, the PZ and PBE results agree with the experiments of Au \cite{lynn} and Cu \cite{nicklow}, respectively. This tendency is similar to the previous calculations \cite{tang}. For AuCu the lowest $\omega$ at point R is sensitive to the functional used: $\hbar\omega = -2.3$ and $1.8$ meV for PZ and PBE, respectively. For CuAu$_3$ the phonon softening at the point M is observed. For AuCu$_3$, the agreement between the calculated and experimental curves \cite{katano} is good when the PBE functional is used. Below we will use the PBE functional to study the dynamical stability of AuCu in detail. 

\begin{figure}[tt]
\center
\includegraphics[scale=0.35]{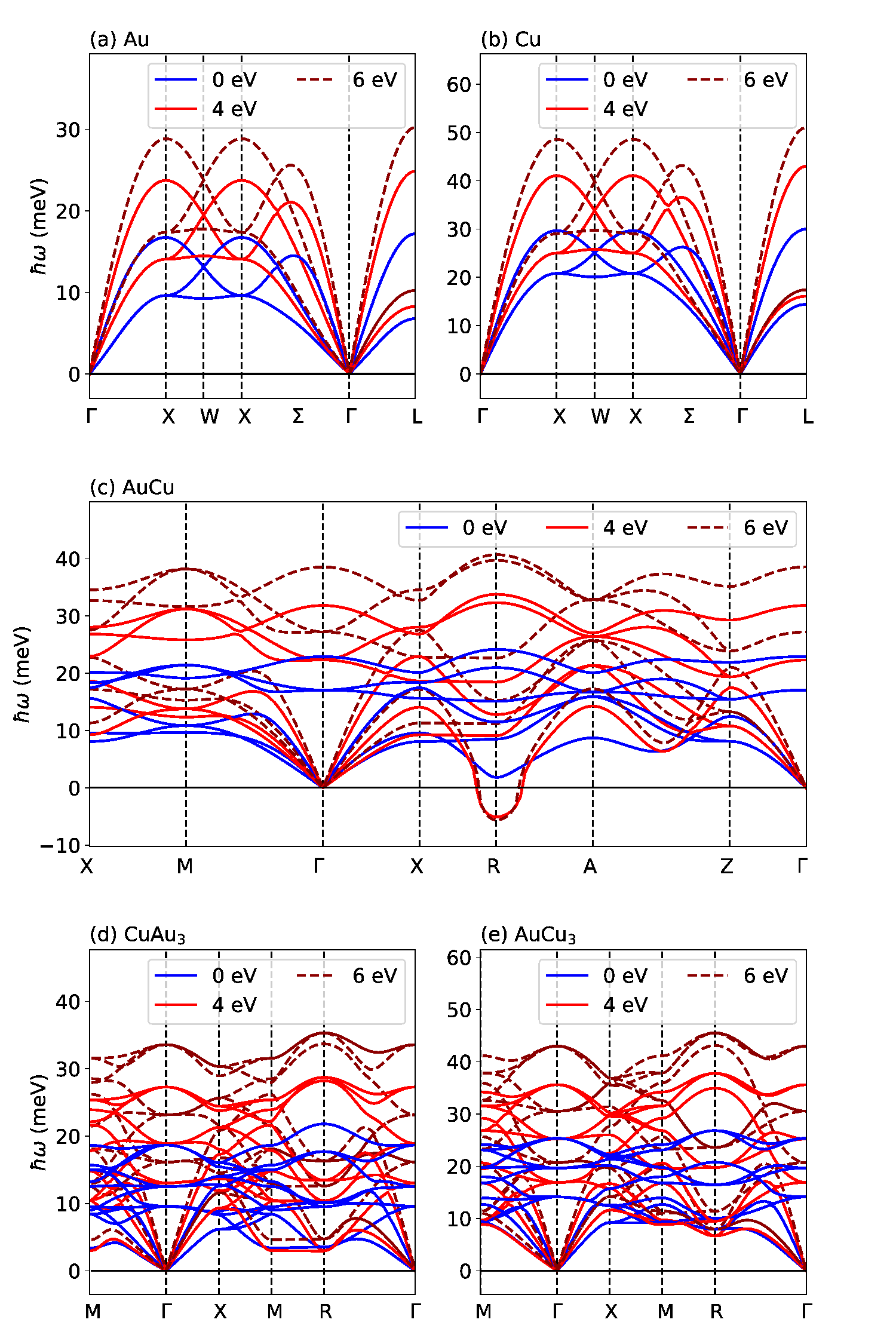}
\caption{The phonon dispersions of (a) Au, (b) Cu, (c) AuCu, (d) CuAu$_3$, and (e) AuCu$_3$ for $k_{\rm B}T_{\rm e}=0, 4,$ and 6 eV. The PBE functional is used. } \label{fig_wdm} 
\end{figure}

\begin{figure}[tt]
\center
\includegraphics[scale=0.5]{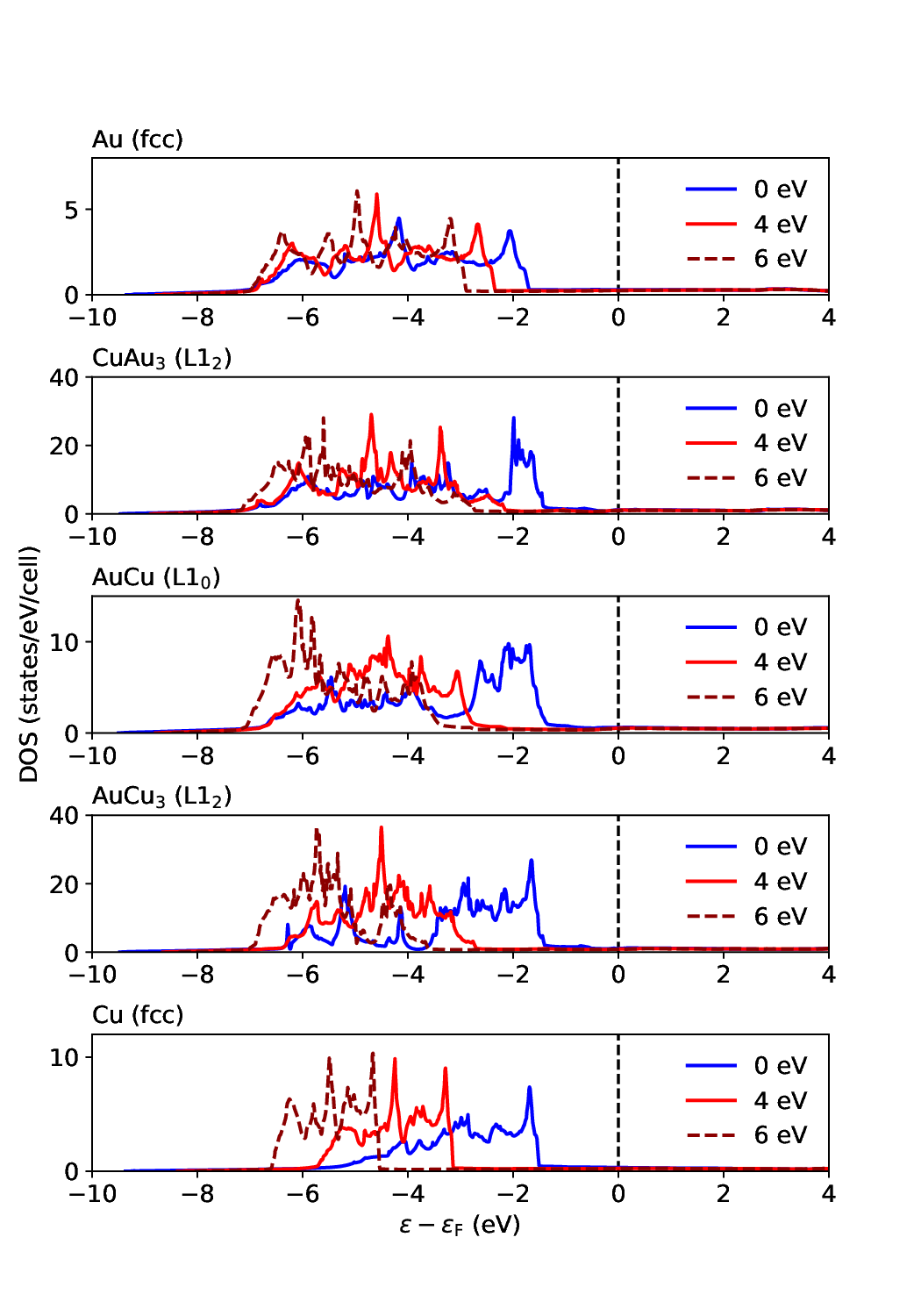}
\caption{The electron DOS for Au, CuAu$_3$, AuCu, AuCu$_3$, and Cu for $k_{\rm B}T_{\rm e}=0, 4$, and 6 eV. The electron energy is measured from the Fermi level.} \label{fig_dos} 
\end{figure}


\subsection{Phonons in the WDM regime}
\label{sec:phonon}
Figure~\ref{fig_wdm} shows the phonon dispersion curves of (a) Au, (b) Cu, (c) AuCu, (d) CuAu$_3$, and (e) AuCu$_3$ in the WDM regime, where $T_{\rm e}$ is set to 0, 4, and 6 eV. For both simple metals (Au and Cu), the phonon energy increases monotonically (i.e., phonon hardening) with $T_{\rm e}$, which is consistent with the previous calculations \cite{recoules,yan}. For CuAu$_3$ and AuCu$_3$ in the L1$_2$ structure, the phonon hardening behaviors are also observed when $T_{\rm e}$ is increased, while nonmonotonic increases are observed along the line of M-R for the transverse acoustic phonon branch. As shown in Fig.~\ref{fig_wdm}(c), L1$_0$ AuCu becomes unstable against $T_{\rm e}$. While the optical phonons are hardened significantly, the transverse acoustic mode is unstable because around point R the phonon energy is imaginary. 




In general, the phonon hardening is attributed to an increase in the internal pressure caused by weakened electron screening. When the screening effect is weakened, bared ions are created, leading to a peak shift and a shrinkage of the width in the electron energy spectrum. This has been illustrated by the electron DOS as a function of $T_{\rm e}$ \cite{recoules}. Figure \ref{fig_dos} shows the DOS of Au-Cu alloys for several $T_{\rm e}$s. At $T_{\rm e}=0$, due to the presence of $d$ bands, the high DOS is observed below the Fermi level from $-2$ to $-7$ eV, while the band width is narrow in Cu (from $-2$ to $-5$ eV). When $T_{\rm e}$ is increased, the peak position and the width become deep and small, respectively, as described above. Although the variation of DOS with $T_{\rm e}$ may be correlated with the phonon hardening of Au, Cu, CuAu$_3$, and AuCu$_3$, it never explains the phonon softening observed in L1$_0$ AuCu. 



\begin{table}[bb]
\begin{center}
\caption{Phonon energies (in units of meV) at point R in L1$_0$ AuCu at $T_{\rm e}=0$ eV for different models. }
{
\begin{tabular}{lcccccc}\hline\hline
   \hspace{5mm} & \hspace{1mm} 1NN \hspace{1mm} & \hspace{1mm} 2NN \hspace{1mm} & \hspace{1mm} 3NN \hspace{1mm}  & \hspace{1mm} 4NN \hspace{1mm} & \hspace{1mm} 5NN \hspace{1mm} & \hspace{1mm} DFPT \hspace{1mm}  \\  \hline
$\omega_{\rm I}$ & -6.1 & -5.6 &	-4.4 &	-3.2 &	0.5  &	1.8  \\
$\omega_{\rm II}$ & 8.0 & 6.4 &	 6.2 &	6.6 &	7.2  &	8.5  \\
$\omega_{\rm III}$ & -6.9 & 9.8 &	10.2 &	10.4 &	10.9 &	11.5 \\
$\omega_{\rm IV}$ & 14.0 & 13.2 &	13.0 &	13.4 &	14.3 &	15.1 \\
$\omega_{\rm V}$ & 19.0 & 19.8 &	19.9 &	20.1 &	20.5 &	21.0 \\
$\omega_{\rm VI}$ & 22.3 & 23.2 &	23.6 &	23.7 &	24.0 &	24.2 \\
\hline\hline
\end{tabular}
}
\label{table_freq1}
\end{center}
\end{table}

\begin{table}[bb]
\begin{center}
\caption{Same as Table~\ref{table_freq1} but for WDM regime at $T_{\rm e}=4$ eV.}
{
\begin{tabular}{lcccccc}\hline\hline
   \hspace{5mm} & \hspace{1mm} 1NN \hspace{1mm} & \hspace{1mm} 2NN \hspace{1mm} & \hspace{1mm} 3NN \hspace{1mm}  & \hspace{1mm} 4NN \hspace{1mm} & \hspace{1mm} 5NN \hspace{1mm} & \hspace{1mm} DFPT \hspace{1mm}  \\  \hline
$\omega_{\rm I}$ & -12.9 & -8.5 &	-5.8 &	-5.7 &	-5.6 &	-5.1 \\
$\omega_{\rm II}$ & 11.0 & 8.3 & 8.1 &	8.2 &	8.3 &	9.1 \\
$\omega_{\rm III}$ & -11.4 & 11.1 &	12.1 &	12.2 &	12.4 &	12.8 \\
$\omega_{\rm IV}$ & 19.3 & 17.6 &	17.5 &	17.6 &	17.8 &	18.5 \\
$\omega_{\rm V}$ & 28.0 & 31.8 &	32.0 &	32.0 &	32.1 &	32.3 \\
$\omega_{\rm VI}$ & 31.9 & 33.0 &	33.6 &	33.7 &	33.7 &	33.8 \\
\hline\hline
\end{tabular}
}
\label{table_freq2}
\end{center}
\end{table}

To understand the anomalous softening of AuCu in the WDM regime, let us remind that the ion-ion interactions are decomposed into the direct and indirect parts: the former is the repulsive Coulomb interaction that is independent of the magnitude of $T_{\rm e}$, while the latter is the electron-mediated attractive interaction that is significantly modified in the WDM regime. At ambient condition, the former and the latter may be partly canceled with each other, producing the repulsive and attractive interactions for the short- and long-range parts, respectively. Given small indirect contribution, the repulsive Coulomb forces are dominant for the short-range part. On the other hand, for the long-range part, the attractive forces compete with the repulsive forces, creating relatively small interaction forces. This situation is exactly realized in WDMs. We thus hypothesize that the lowest energy mode at point R of L1$_0$ AuCu is stabilized by the long-range interaction between atoms at $T_{\rm e}=0$ eV. With this assumption, it is reasonable to observe the softening of the phonon mode when L1$_0$ AuCu is in the WDM regime having weakened long-range interatomic interactions. Similarly, we can expect that when a phonon mode is stabilized by only the short-range interactions, such a mode will show hardening behavior in the WDM regime.  


To demonstrate this concept, we list the values of $\omega_k$ in Eq.~(\ref{eq:omegaI-VI}) for several models accounting for up to the $p$-th NN interactions ($p$NN models) in Table \ref{table_freq1}. As $p$ increases, the phonons at point R are stabilized and their energies approach the DFPT results. It is noteworthy that the sizes of $\omega_{\rm V}$ and $\omega_{\rm VI}$ within the 1NN model are almost comparable to those within DFPT. This means that the optical phonon energies are well determined by only the short-range interaction between atoms. When the interatomic interactions are considered up to the second order NN, $\omega_{\rm III}$ becomes positive. However, $\omega_{\rm I}$ is positive only within 5NN and DFPT. This implies that the long-range forces between Au and Cu atoms are important to yield positive $\omega_{\rm I}$, i.e., the dynamical stability of L1$_0$ AuCu. Table \ref{table_freq2} also lists the values of $\omega_k$ but for $T_{\rm e}=4$ eV. As in Table \ref{table_freq1}, $\omega_k$ approaches the DFPT results with the inclusion of higher order NN interactions. Again, the sizes of $\omega_{\rm V}$ and $\omega_{\rm VI}$ within the 1NN model almost agree with those within DFPT, indicating that the optical phonon hardening is due to an enhanced short-range forces. The most important fact is that the convergence of $\omega_{\rm I}$ is fast compared to the case of $T_{\rm e}=0$ eV: $\omega_{\rm I}=-5.8$ meV within 3NN is already similar to $-5.1$ meV within DFPT. This reflects the short-range nature of WDM, leading to the instability of the $\omega_{\rm I}$ mode. 

\begin{figure}[tt]
\center
\includegraphics[scale=0.35]{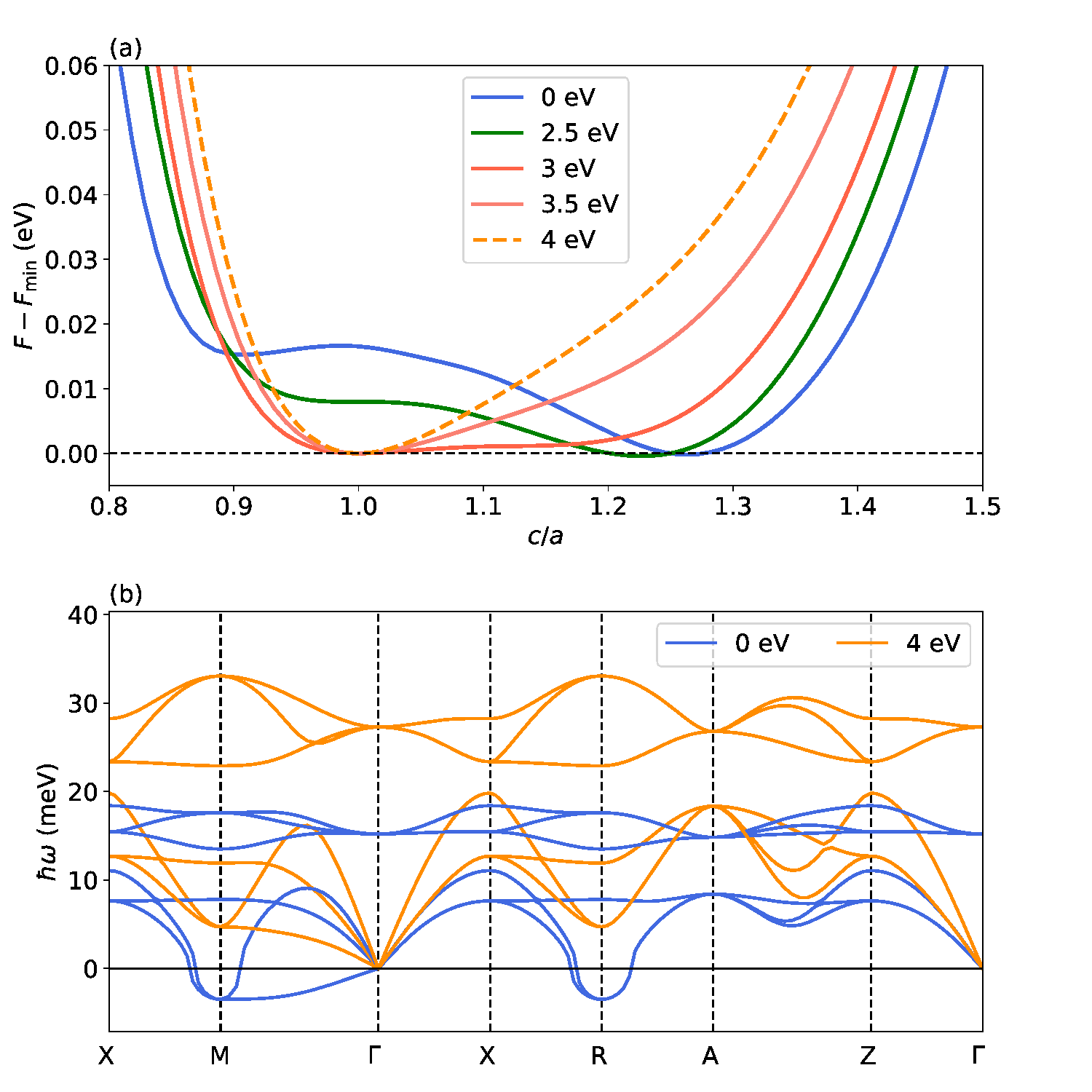}
\caption{(a) The free-energy of L1$_0$ AuCu along the Bain path for different $T_{\rm e}$s. The free-energy is measured from the minimum value along $c/a$ with $T_{\rm e}$ fixed. (b) The phonon dispersion curves of AuCu with $c/a=1$ (i.e., B2) for $T_{\rm e}=0$ and $4$ eV. } \label{fig_bain} 
\end{figure}

\subsection{Solid-to-solid transformation}

The stability of AuCu in the L1$_0$ structure can also be understood by calculating the free-energy as a function of the ratio $c/a$ with the equilibrium volume $V=V_0$ fixed (i.e., along the Bain path \cite{grimvall}): $c/a=1$ and $\sqrt{2}$ correspond to the bcc (or B2) and fcc structures, respectively. Figure \ref{fig_bain}(a) shows the $c/a$-dependence of the free-energy of AuCu for several $T_{\rm e}$s, where $V=V_0$ is the equilibrium volume of L1$_0$ at $T_{\rm e}=0$ eV. At $T_{\rm e}=0$ eV, the free energy takes minimum values at $c/a=1.26$ (L1$_0$) and $0.9$, whereas that takes a maximum value at $c/a=1$ (B2). As expected, AuCu in the B2 structure is dynamically unstable (see Fig.~\ref{fig_bain}(b) (blue)). When $T_{\rm e}$ is increased, the profile of such a double-well-like curve is modified, indicating a non-thermal solid-to-solid transformation, as discussed in warm dense tungsten \cite{giret}. At $T_{\rm e}=2.5$ eV, the free-energy minimum at smaller $c/a$ vanishes, while that at larger $c/a$ shifts to $c/a\simeq 1.22$; at $T_{\rm e}=3$ eV, the free-energy curve becomes relatively flat in the region of $c/a\in [1.0,1.2]$ and takes a minimum only at $c/a=1$; and at higher $T_{\rm e}$s, the curvature around $c/a=1$ becomes larger, stabilizing the B2 phase. Such a B2 structure is found to be dynamically stable, as shown in Fig.~\ref{fig_bain}(b) (orange), where $T_{\rm e}=4$ eV is assumed. We hope that our prediction for the phase transformation in AuCu, before melting occurs, can be confirmed by future experiments.  




\section{Conclusions} 
\label{sec:conclusion}
In conclusion, we have studied the dynamical stability of Au-Cu alloys in the WDM regime by calculating the phonon dispersion curves. The phonon hardening is observed in the L1$_2$ and L1$_0$ structures except the lowest frequency phonon at point R of L1$_0$ AuCu. The hardening behavior can be explained by the increase in the internal pressure caused by the $d$ electron excitations, as observed in noble metals. The phonon softening observed is attributed to the short-range nature of WDMs because the lowest frequency phonon at point R is stabilized by long-range interactions between Au and Cu. 

Although we have used analytical expressions of Eq.~(\ref{eq:omegaI-VI}) to understand the phonon softening of L1$_0$ AuCu, those can be applicable to other ordered alloys in the L1$_0$ and B2 (the case of $c=a$) structures. It is valuable to note that 51 L1$_0$ and 325 B2 alloys are found in experimental phase diagrams \cite{sluiter}. More investigations will lead to a deep understanding of the dynamical stability of ordered alloys. 



Finally, it should be noted that the WDM is a transient phase because the excited electron energy created by a laser pulse is transferred to phonons, leading to lattice disordering or melting within a few picoseconds \cite{cho,jourdain2018,mo,dara,smirnov,jourdain2021}. Although the phonon hardening of warm dense Au has been reported experimentally \cite{ernstorfer}, its validity has been controversial \cite{dara,smirnov,jourdain2021}. It is thus desirable to simulate the time-evolution of lattice displacements in warm dense alloys accurately, which enables us to interpret time-resolved experiments.  

\begin{acknowledgments}
A part of numerical calculations has been done using the facilities of the Supercomputer Center, the Institute for Solid State Physics, the University of Tokyo.
\end{acknowledgments}

\appendix

\section{Dynamical matrix}
\label{app}
The expressions of $\tilde{D}_{\alpha\beta}^{\kappa\kappa'}(\bm{q})$ used in the $p$NN model calculations are given below. The symmetry properties for the force constants \cite{maradudin} are used to simplify the expression of Eq.~(\ref{eq:dyn}). For the L1$_0$ structure, there are 16 symmetry operations including the identity, 180 degree rotation around $x, y, z$, and $[1,\pm 1,0]$ axis, $\pm$90 degree rotation around $z$ axis, and the inversions. The list of the force constants obtained from DFPT calculations is used to check the restrictions imposed. We introduce the step function $U(x)$ that takes $1$ and $0$ for $x\ge 0$ and $x< 0$, respectively. We obtain for $\alpha=x$ and $y$,
\begin{eqnarray}
 & &  M_\kappa \tilde{D}_{\alpha\alpha}^{\kappa\kappa}(\bm{q}) 
 \nonumber\\
 &=& D_{\alpha\alpha}^{\kappa\kappa}(\bm{0}) U(p-1)
 \nonumber\\
 &+& 2\left[ D_{\alpha\alpha}^{\kappa\kappa}(\bm{a}_1) - D_{\alpha\alpha}^{\kappa\kappa}(\bm{a}_1) - D_{\alpha\alpha}^{\kappa\kappa}(\bm{a}_1)\right]  U(p-2)
 \nonumber\\
 &+& 4\left[-D_{\alpha\alpha}^{\kappa\kappa}(\bm{a}_1 + \bm{a}_3) + D_{\alpha\alpha}^{\kappa\kappa}(\bm{a}_2+\bm{a}_3) \right]  U(p-4),
 \nonumber\\
 \label{eq:Dxxyy}
\end{eqnarray}
and for $\alpha=z$,
\begin{eqnarray}
 M_\kappa \tilde{D}_{\alpha\alpha}^{\kappa\kappa}(\bm{q}) 
 &=& D_{\alpha\alpha}^{\kappa\kappa}(\bm{0}) U(p-1)
  \nonumber\\
 &-& 2 D_{\alpha\alpha}^{\kappa\kappa}(\bm{a}_3) U(p-3). 
 \label{eq:Dzz}
\end{eqnarray}
The nonzero elements in the off-diagonal matrix for $(\alpha\beta)=(yz)$ and $(zy)$ are
\begin{eqnarray}
& & \sqrt{M_1 M_2} \tilde{D}_{\alpha\beta}^{12}(\bm{q}) 
\nonumber\\
 &=& 8 D_{\alpha\beta}^{12}(\bm{0}) U(p-1)
 \nonumber\\
 &+& 8 \left[ 
   D_{\alpha\beta}^{12}(2\bm{a}_1) 
+ D_{\alpha\beta}^{12}(2\bm{a}_2)   
 \right] U(p-5).
\end{eqnarray}
The dynamical matrix is symmetric because of the relation $\tilde{D}_{\beta\alpha}^{21}(\bm{q})=\tilde{D}_{\alpha\beta}^{12}(\bm{q})$. The acoustic sum rule determines $D_{\alpha\alpha}^{\kappa\kappa}(\bm{0})$ in Eqs.~(\ref{eq:Dxxyy}) and (\ref{eq:Dzz}), depending on the model used.


\end{document}